# In-plane Néel wall chirality and orientation of interfacial Dzyaloshinskii-Moriya vector in magnetic films

MacCallum Robertson[1], Christopher J. Agostino[2,3,4], Gong Chen[1,*], Sang Pyo Kang[5], Arantzazu Mascaraque[6,7], Enrique Garcia Michel[8], Changyeon Won[5], Yizheng Wu[9], Andreas K. Schmid[2], and Kai Liu[1,10,*]

[1]Department of Physics, University of California, Davis, California 95616, USA
[2]NCEM, Molecular Foundry, Lawrence Berkeley National Laboratory, Berkeley, California 94720, USA
[3]Physics Department, University of California, Berkeley, California 94720, USA
[4]Department of Astronomy, Indiana University, Bloomington, Indiana 47405, USA
[5]Department of Physics, Kyung Hee University, Seoul 02447, Korea
[6]Dpto. de Física de Materiales, Universidad Complutense de Madrid, 28040 Madrid, Spain
[7]Unidad Asociada IQFR(CSIC)-UCM, Madrid E-28040, Spain
[8]Dept. Condensed Matter Physics, and IFIMAC-Condensed Matter Physics Center, Universidad Autonoma de Madrid, 28049 Madrid, Spain
[9]Department of Physics, State Key Laboratory of Surface Physics, and Advanced Materials Laboratory, Fudan University, Shanghai 200433, China
[10]Physics Department, Georgetown University, Washington, DC 20057, USA
* Correspondence should be addressed to gchenncem@gmail.com (G.C.); Kai.Liu@georgetown.edu (K.L.)

## Abstract

The interfacial Dzyaloshinskii-Moriya interaction (DMI) is of great interest as it can stabilize chiral spin structures in thin films. Experiments verifying the orientation of the interfacial DMI vector remain rare, in part due to the difficulty of separating vector components of DMI. In this study, Fe/Ni bilayers and Co/Ni multilayers were deposited epitaxially onto Cu(001) and Pt(111) substrates, respectively. By tailoring the effective anisotropy, spin reorientation transitions (SRTs) are employed to probe the orientation of the DMI vector by measuring the spin structure of domain walls on both sides of the SRTs. The interfacial DMI is found to be sufficiently strong to stabilize chiral Néel walls in the out-of-plane magnetized regimes, while achiral Néel walls are observed in the in-plane magnetized regimes. These findings experimentally confirm that the out-of-plane component of the DMI vector is insignificant in these fcc(001) and fcc(111) oriented interfaces, even in the presence of atomic steps.

**Introduction**

Spin textures such as magnetic domain walls (DWs), spin spirals, and magnetic skyrmions have emerged as an exciting playground for explorations of topological characteristics of nanomagnetic elements and spin-orbit coupling at surfaces and interfaces [1-6]. They have important potential applications in high density, low energy consumption spintronic devices [6-8]. The nucleation and stabilization of such spin textures usually depends on the interplay between the exchange interaction, magnetic dipolar interaction, magnetic anisotropy, and the Dzyaloshinskii-Moriya interaction (DMI). The non-collinear components of spin textures such as domain walls are usually expected to be achiral, where right-handed and left-handed rotating spin structures are energetically degenerate as a result of the symmetric exchange interaction [9]. However, the broken inversion symmetry at interfaces introduces an antisymmetric exchange interaction, the DMI [10,11], which lifts the left/right-handed degeneracy and stabilizes magnetic chirality. The DMI term is written as $E_{\mathrm{DM}} = -\mathbf{D}_{ij} \cdot (\mathbf{S}_i \times \mathbf{S}_j)$ where $\mathbf{D}_{ij}$ is the DMI vector and $\mathbf{S}_i$ and $\mathbf{S}_j$ are the spins on neighboring atomic sites $\mathbf{r}_i$ and $\mathbf{r}_j$. At interfaces in otherwise ideal centrosymmetric lattices, such as fcc(001) and fcc(111) interfaces, the direction of the DMI vector $\mathbf{D}_{ij}$ is predicted to lie in the interface plane and perpendicular to distance vector $\mathbf{r}_{ij} = \mathbf{r}_i - \mathbf{r}_j$ for basic symmetry reasons [11,12].

Chiral spin textures of DWs [2-4] and skyrmions [5,6] are commonly observed in out-of-plane magnetized thin film systems. Here two DW types, Bloch and Néel walls, are characterized by the helical and cycloidal rotation of the spin vector about the DW normal, respectively. The chirality can be stabilized in Néel walls in the presence of significant interfacial DMI [13]: in this case $\mathbf{S}_i \times \mathbf{S}_j$ is collinear with $\mathbf{D}_{ij}$, leading to a non-zero $E_{\mathrm{DM}}$. In contrast, within Bloch walls the helical spin vector rotation generally results in the term $(\mathbf{S}_i \times \mathbf{S}_j)$ being collinear with $\mathbf{r}_{ij}$ and orthogonal to $\mathbf{D}_{ij}$, leading to a zero $E_{\mathrm{DM}}$ and achiral Bloch walls [2].

In in-plane magnetized systems the observation of DMI-stabilized chirality is much less common [8]. At ideal interfaces between two layers, both $\mathbf{D}_{ij}$ and the spins $\mathbf{S}_i$ and $\mathbf{S}_j$ are lying perfectly within the interface plane resulting in a zero $E_{\mathrm{DM}}$ [12]. However, non-zero out-of-plane DMI vector may appear in specific atomic configurations near a bcc(111) or fcc(110) surface [12], or in case of interlayer Dzyaloshinskii-Moriya Interactions [14]. Moreover, atomic scale roughness, vicinal interface orientation, and other symmetry-reducing imperfections can cause canting such that the orientation of spins [15,16] as well as $\mathbf{D}_{ij}$ may acquire certain out-of-plane components, which can lead to finite values of $E_{\mathrm{DM}}$. The distinction between the ideal and realistic cases is important - multiple studies have found a finite DMI in symmetric systems such as Pt/Co/Pt trilayers [17] or Co/Pd multilayers [18], which are expected to exhibit zero DMI. Additionally, in certain magnetic nanodisks in the vortex state, a preferred circularity of the vortex with a given polarity of the vortex core has been attributed to the DMI at the surface of the nanodisks [19]. In some magnetic stripe-shaped nanoislands, the magnetization at the two ends of the islands is found to tilt towards the out-of-plane direction with preferred chirality due to the interfacial DMI [20]. In certain thin films, magnetic chirality has been observed in unusual out-of-plane domain walls due to the interplay between uniaxial strain, effective anisotropy, and the DMI [21]. More recently, chiral coupling between controlled in-plane and out-of-plane magnetized regions was realized in special anisotropy engineered nanomagnets [22]. The stabilization of magnetic chirality in these in-plane magnetized systems involves out-of-plane components of magnetization, while the study of magnetic chirality in purely in-plane magnetized thin films and the potential influence of a finite out-of-plane component of the DMI vector $\mathbf{D}_{ij}$ remain rare.

In the present work we investigate the rotation sense of in-plane Néel walls in purely in-plane magnetized systems with significant DMI [23-26]. We first study spin structures within domain walls in out-of-plane magnetized systems of two different fcc surface orientations to confirm that the interfacial DMI is strong enough to stabilize chiral Néel walls. In the same systems we then facilitate spin-reorientation transitions (SRTs) to turn their magnetization in-plane, without changing the DMI interfaces, and investigate the chirality of the resulting in-plane Néel walls. We demonstrate that upon SRT the domain wall structures in the in-plane magnetized regimes are achiral in-plane Néel walls. This can be understood as an experimental confirmation of the in-plane orientation of the DMI vector, i.e. possible out-of-plane canting of $\mathbf{D}_{ij}$ is negligible. Finally, we comment on how the chiral degeneracy of in-plane Néel walls in such systems may be lifted.

**Experimental Methods**

The experiments were performed in the spin-polarized low-energy electron microscope (SPLEEM) at the National Center for Electron Microscopy of the Lawrence Berkeley National Laboratory [27,28]. A GaAs-type spin polarized electron gun and spin manipulator enables alignment of the electron beam spin polarization in any direction with angular resolution of ~1° [29,30]. Magnetic contrast along cartesian in-plane and out-of-plane directions enables real space mapping of the 3-dimensional magnetization vector $M_x$, $M_y$, and $M_z$ [31]. The high spatial-resolution of SPLEEM images allows nanoscale mapping of the spin configurations within the DWs, which enables determination of the DW chirality [32]. The incident electron energy is selected to be 5eV to optimize magnetic contrast. For sample preparation, a Pt(111) substrate was cleaned by several cycles of flashing at 850 °C in 3x10$^{-8}$ torr O$_2$ and again at 1100 °C in 4x10$^{-11}$ torr vacuum. A Cu(001) substrate was cleaned by cycles of Ar$^+$ sputtering at 1keV and annealing at 600 °C until signals of possible contaminants including carbon, sulfur and oxygen are no longer detectable in Auger electron spectra. Fe/Ni and Co/Ni layers were deposited on Cu(001) and Pt(111), respectively, at room temperature via electron beam evaporation at a base pressure below $5 \times 10^{-11}$ torr. By monitoring oscillations in the low energy electron microscopy (LEEM) intensity associated with layer-by-layer growth, monolayer (ML) control of the film thickness was achieved. The evaporators were positioned facing the substrate at a grazing angle of 15° with respect to the sample surface. SPLEEM images were taken at room temperature, generated from the polarization-dependent reflectivities of the spin-up and spin-down electrons $I_\uparrow$ and $I_\downarrow$, respectively, where the pixel-by-pixel magnetic contrast is $A = \frac{I_\uparrow - I_\downarrow}{I_\uparrow + I_\downarrow}$. In conjunction with the experimental data, we used Monte Carlo simulations based on a two-dimensional model that includes the exchange interaction, magnetic anisotropy, dipolar interaction, and interfacial DMI to model the magnetic domain walls [33].

**Results and Discussion**

We selected two systems, (i) [Co/Ni]$_n$ multilayer stacks on Pt(111) and (ii) Fe/Ni bilayer stacks on Cu(001), as both systems exhibit very sensitive thickness-dependent SRTs. Specifically, in the [Co/Ni]$_n$/Pt (111) system the easy axis of magnetization rotates from out-of-plane to in-plane orientation with the addition of a single monolayer of Co; Similarly the magnetic configuration of Fe/Ni bilayers grown on Cu(001) changes from an out-of-plane magnetized stripe phase to an in-plane magnetized phase when the thickness of the Fe layer is increased from 2ML to 3ML [34,35]. The significance of selecting systems with such clearly defined SRTs is that this allows us to first test for presence of significant interfacial DMI by measuring the chirality of magnetic domain walls while the systems are in their out-of-plane magnetized states and subsequently drive the systems through their SRTs, by slightly increasing film thickness, to observe effects of the DMI on DW chirality in in-plane magnetized samples.

We first discuss the out-of-plane magnetized fcc(001) system, Fe (2.5ML)/Ni (2ML) on Cu (001). Fig. 1(a) shows a sketch of the sample structure and a corresponding colorized SPLEEM image where the color wheel in the inset indicates the in-plane magnetization direction within the domain walls. The direction of the magnetization vector within these domain walls always points from gray (down domains, negative $M_z$) to black (up domains, positive $M_z$) regions, suggesting that the DWs are chiral [2]. To quantify this chirality, we define the angle $\alpha$ between the magnetization vector $\boldsymbol{m}$ and the DW normal vector $\boldsymbol{n}$ at each pixel in the DW, pointing from gray to black domains, as sketched in the inset. The magnitude of the angle $\alpha$ indicates whether DW pixels are Bloch type ($\alpha = \pm 90°$), Néel type ($\alpha = 0°$ or $180°$), or mixed [36]. A histogram of the angle $\alpha$ measured from a large ensemble of DWs then illustrates the degree of domain wall chirality. For example, two peaks at $\alpha = 0°$ and $\alpha = 180°$ would indicate achiral Néel DWs, while histograms with a single peak at either $\alpha = 0°$ or $\alpha = 180°$ would indicate chiral Néel DWs. The $\alpha$ histogram of this sample shows a single peak centered around $\alpha = 0°$ [Fig. 1(c)], indicating that the out-of-plane magnetic domains in this sample are separated by chiral (left-handed) Néel DWs as a result of the interfacial DMI contribution to the system. We then deposit an additional 1ML of Fe to the stack to induce an SRT to an in-plane easy magnetization axis. Both 180° and 90° domain walls (not shown here) are observed in in-plane magnetized region. However, the degeneracy of opposite in-plane rotation senses of 90° domain walls is lifted due to the different exchange energy penalties between 90° and 270° magnetization rotation, which is related to the magnetization direction of adjacent domains. On the other hand, the exchange energy penalties of 180° domain walls with opposite in-plane rotation senses remain the same, which is ideal to test the role of the DMI on the chirality. The resulting magnetic microstructure of 180° domain wall is illustrated in Fig. 1(b), where magnetization orientation of the in-plane domains is given by the corresponding color wheel. To quantify the chirality in the domain wall, we define an angle $\chi$ as the angle between the magnetization vector $\boldsymbol{m}$ and the domain magnetization vector $\boldsymbol{m_d}$ at each pixel in the domain wall [see inset in Fig. 1(d)]. The $\chi$ histogram of this Fe(3.5ML)/Ni (2ML)/Cu (001) sample exhibits two peaks around 90° and -90° (270°), as shown in Fig. 1(d), indicative of the presence of similar populations of DWs with both in-plane magnetization rotation senses (clockwise and counterclockwise). Thus this sample has an achiral Néel DW structure. Note that the DW structures in out-of-plane and in-plane magnetized systems are both called Néel-type DW because of their cycloidal type rotation [8,37]. The presence of the anisotropic environments such as uniaxial anisotropy [38-40] or anisotropic DMI [41] in in-plane magnetized system is also expected to stabilize achiral Néel wall as the 180° domain wall is preferred, and the chirality may only occur in out-of-plane domain walls when the anisotropic term dominates [21]. Both the out-of-plane and the in-plane Fe/Ni/Cu (001) samples have identical interfaces, therefore the possibility that the absence of DW chirality in the in-plane magnetized films might be due to insufficient interfacial DMI can be ruled out (in fact, the strength of the DMI in this system was measured previously [2] to be 0.12-0.17 meV per atom).

To investigate a system with a different crystal orientation – and an even larger value of the DMI – we now discuss measurements on fcc(111) [Co/Ni]$_n$ multilayer stacks on Pt(111). Fig. 2(a) shows how, for a Ni(2ML)/Co (1ML)/Ni (2ML)/Pt (111) sample, the magnetization within the domain walls points from gray to black regions, which clearly suggests that the system is chiral. For this sample the histogram of the angle $\alpha$, as defined earlier, has a peak centered around $\alpha = 0°$, indicating the presence of chiral (left-handed) Néel walls. We then deposited an additional 1ML of Co onto this stack in order to induce the SRT to in-plane magnetization, as shown in Fig. 2(b). It has been shown in previous studies that the DMI at this additional Co/Ni interface is negligible compared to the overall DMI of the system [42], and that the DMI strength at the Ni/Pt interface is sufficient to stabilize magnetic chirality even in much thicker films [36]. The corresponding $\chi$ histogram of this in-plane magnetized sample exhibits the characteristic double peaks at $\chi = \pm 90°$ [Fig. 2(d)] associated with achiral Néel walls. Again, the strong DMI present in this system is sufficient to stabilize chiral Néel walls in the out-of-plane magnetized case, while the in-plane magnetized counterpart only hosts achiral Néel walls.

These results are further corroborated by our micromagnetic simulations based on a two-dimensional model described in refs [43,44] where exchange interaction $J$, dipolar interaction $D_{\text{Dip}}$, effective magnetic anisotropy $K_z^{\text{eff}}$, as well as the DMI $\mathbf{D}_{ij}$ are considered [2,21,36]. In Fig. 3(a) an out-of-plane magnetized domain configuration is simulated with dimensionless parameters $J = 1$, $D_{\text{Dip}} = 0.1$, $D_{ij} = 0.1$, $K_z^{\text{eff}} = 0.03$ (positive/negative $K_z^{\text{eff}}$ corresponds to out-of-plane/in-plane easy axis of magnetization), which reproduces the experimental domain structures with chiral Néel walls shown in Figs. 1(a) and 2(a). Next we set $K_z^{\text{eff}}$ to -0.43 to simulate in-plane anisotropy, while keeping the other parameters unchanged. We find that domain configuration evolves to in-plane magnetized domains separated by Néel walls with no preferred rotation sense, as indicated by the presence of both green/yellow and blue/magenta walls along the domain boundary in Fig. 3(b). Additionally, Bloch lines appear in in-plane achiral magnetized walls in Fig. 3(b), denoted by black and white dots along the domain wall. Such Bloch lines are expected to be achiral due to the vanishing DMI energy of the Bloch/helical-type domain/domain wall structures surrounding the Bloch lines, which is supported by the observation of achiral Bloch-components in domain walls in both in-plane and out-of-plane magnetized systems [21,45]. These simulation results reproduce the achiral nature of the Néel walls observed experimentally in Figs. 1(b) and 2(b).

Observing chiral Néel walls in the fcc(111) and fcc(001) out-of-plane magnetized films demonstrates the presence of significant interfacial DMI in these systems. After driving the systems through spin reorientation transitions to in-plane magnetized states we find that the DMI is insufficient to break the degeneracy between the two possible rotation senses (clockwise and counterclockwise) of their Néel domain walls. In the following section, we describe in more detail how this chiral-to-achiral transition accompanying the SRT represents experimental evidence supporting the idea that the orientation of the DMI vector does not significantly cant away from the film plane. We also discuss conditions that might support a larger normal component of the DMI vector and possibly the stabilization of chiral Néel walls separating in-plane domains.

For ideal interfaces in fcc(111) and fcc(001) lattices, the DMI vector is predicted to lie in the plane as sketched in Figs. 4(a) and 4(b) , respectively [11,12]. For Néel walls in out-of-plane magnetized systems, the cross product of $\mathbf{S}_i \times \mathbf{S}_j$ is a vector that also lies in the plane, as shown in Fig. 4(c). This orientation results in a nonzero value for $E_{\mathrm{DM}} = -\mathbf{D}_{ij} \cdot (\mathbf{S}_i \times \mathbf{S}_j)$ and thus raises/lowers the free energy of right/left-handed spin textures, which allows for the stabilization of chiral Néel walls, as observed in our out-of-plane magnetized systems. However, the easy axis of magnetization in an otherwise identical system lies within the film plane, therefore inside a Néel wall separating two in-plane domains the $\mathbf{S}_i \times \mathbf{S}_j$ vector is perpendicular to the vector $\mathbf{D}_{ij}$, as shown in Fig. 4(d). This condition results in $E_{\mathrm{DM}} = 0$, i.e. this interfacial DMI does not stabilize a preferred chirality in this type of Néel wall, as we observed in the two in-plane systems studied here.

The absence of chirality in these Néel walls indicates that in the systems [Co/Ni]$_n$/Pt(111) and Fe/Ni/Cu(001) the DMI vector $\mathbf{D}_{ij}$ does not have any significant out-of-plane component. In turn, this interpretation suggests possible paths to stabilize chiral Néel walls in an in-plane magnetized system by amplifying a possible out-of-plane component of the DMI vector. The presence of surface roughness/steps – for example the dark lines in the LEEM image in Fig. 5(a) of our clean Cu(001) surface are surface steps of various heights – may introduce a component of the DMI vector pointing normal to the film surface, as sketched in Fig. 5(b). We have chosen to focus on the out-of-plane component of the DMI vector as it may stabilize chirality in Néel walls between in-plane domains. This is due to the fact that the cross product $\mathbf{S}_i \times \mathbf{S}_j$ of neighboring spin sites along the atomic step, as illustrated in 5(c), is a vector that is collinear with the DMI vector, resulting in a non-zero $E_{\mathrm{DM}}$. Although the array of surface steps on the Cu(001) substrate, as seen in Fig. 5(a), indicates that a finite degree of vicinality is present in our samples, its effect on the DMI vector is evidently not strong enough to influence the chirality [46]. We propose the use of substrates with a significantly larger density of atomic steps, as in a substrate with a strongly vicinal surfaces or any curved surfaces [47,48] to stabilize chiral Néel walls in in-plane magnetized systems.

**Conclusions**

In summary, we have investigated domain wall chirality using SPLEEM in fcc(111) and fcc(001) multilayers with strong DMI. Co/Ni multilayers and Fe/Ni bilayers deposited on Pt(111) and Cu(001) substrates, respectively, provide model systems to explore the chiral nature of the domain walls in out-of-plane *vs.* in-plane magnetized films, before and after they undergo thickness dependent SRTs. For both the fcc(111) and fcc(001) systems chiral Néel walls were observed in the out-of-plane magnetized condition, while in the corresponding in-plane magnetized condition, post-SRT, both systems yielded achiral Néel walls. These findings indicate that in both fcc(111) and fcc(001) interface orientations the DMI vector is oriented completely within the film plane, with no significant out-of-plane component, even in the presence of the a small degree of vicinality seen in the substrates used here. Further experiments are required to test whether the use of substrates with a substantial degree of vicinal miscut may provide a route towards stabilizing chiral Néel walls in in-plane magnetized system.

**Acknowledgement**

This work has been supported by the NSF (DMR-1610060 and DMR-1905468) and the UC Office of the President Multicampus Research Programs and Initiatives (MRP-17-454963). Work at the Molecular Foundry was supported by the Office of Science, Office of Basic Energy Sciences, of the US Department of Energy under contract no. DE-AC02-05CH11231. Work at GU has been supported in part by SMART (2018-NE-2861), one of seven centers of nCORE, a Semiconductor Research Corporation program, sponsored by National Institute of Standards and Technology (NIST). Work at the Kyung Hee University been supported by the National Research Foundation of Korea funded by the Korean Government (NRF-2018R1D1A1B07047114). A. M. acknowledges support from the MINECO under the project MAT2017-87072-C4-2-P. E.G.M acknowledges support from MICINN (Spain) under Grant FIS2017-82415-R and from MECD (Spain) under Grant PRX17/00557. Y.W acknowledges support from National Natural Science Foundation of China (Grant No. 11974079 and 11734006).

**Figures:**

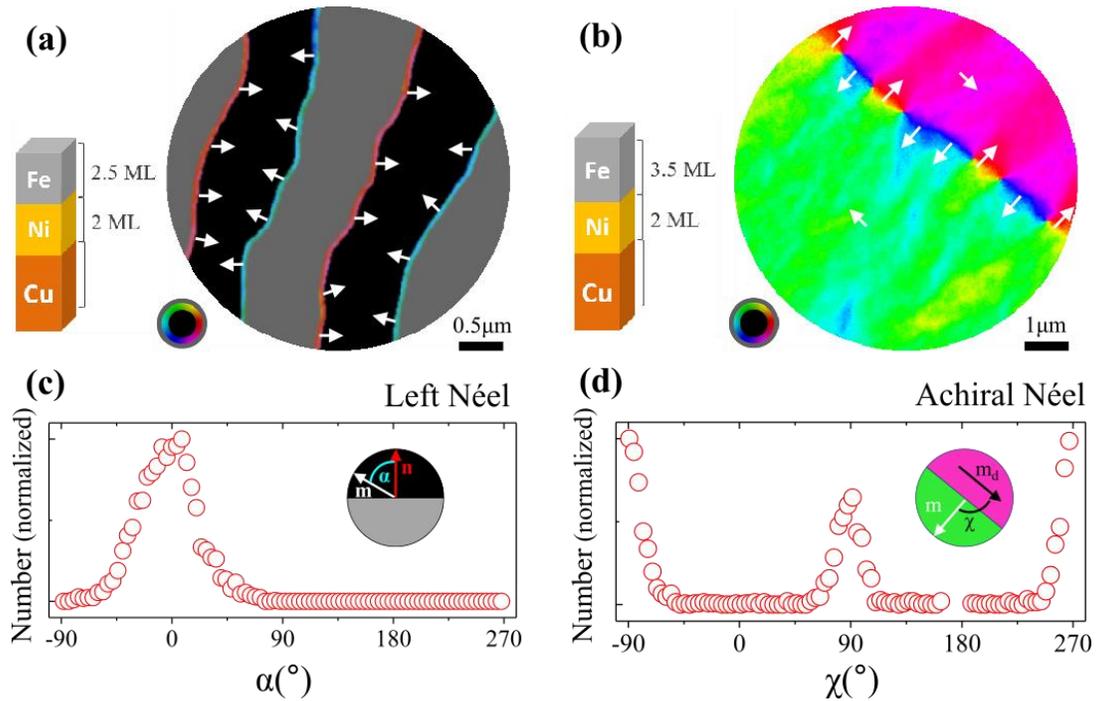

FIG. 1: (a,b) Sketches of sample structures and colorized SPLEEM images for two Fe/Ni/Cu(001) bilayer films, showing the orientation of the local magnetization vector based on the corresponding color wheel in the inset. White arrows additionally highlight domain wall magnetization orientation. (c) Histogram corresponding to (a) showing the normalized counts (each dot representing counts after binning) of the angle α, defined as the angle between the DW magnetization vector $m$ and the DW normal vector $n$, as shown in the inset. A single peak at α=0° demonstrates left-handed Néel DW structure. (d) Histogram corresponding to (b) showing the normalized count of the angle χ, defined as the angle between the DW magnetization vector $m$ and the domain magnetization $m_d$, as shown in the inset; Peaks at χ=±90° and 270° indicate achiral Néel DW structure.

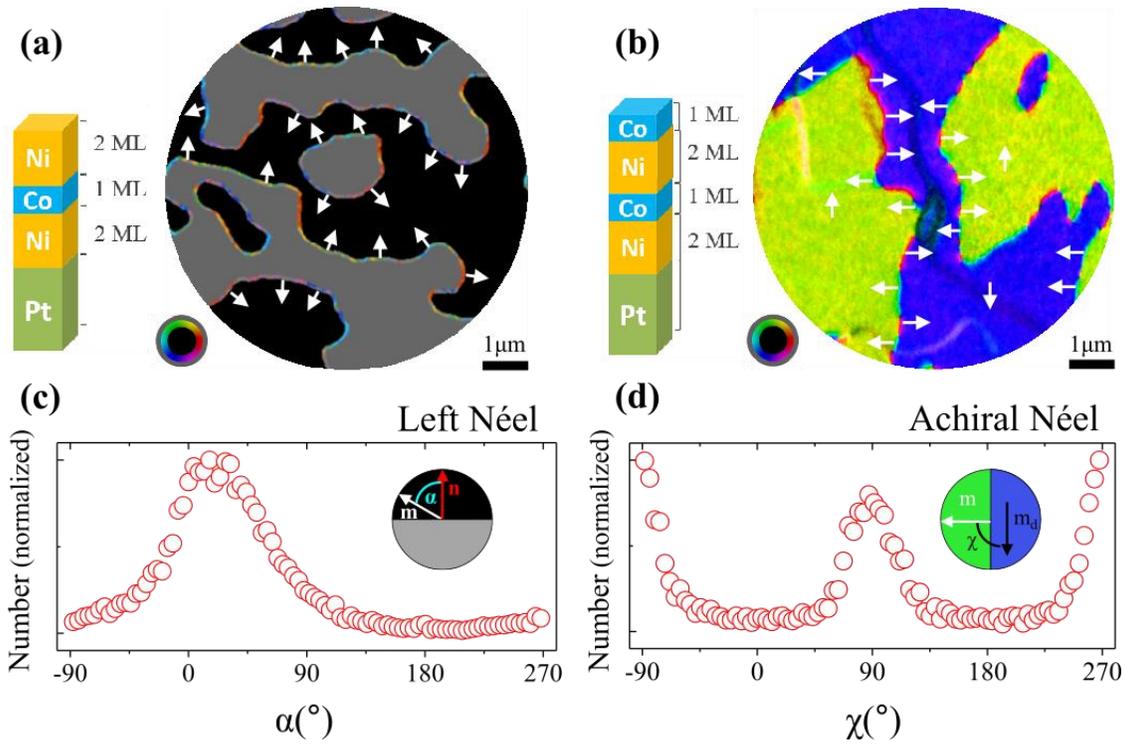

FIG. 2: (a,b) Sketches of sample structures and colorized SPLEEM images for the $[Co/Ni]_n$ on Pt(111) system, showing the orientation of the local magnetization vector based on the corresponding color wheel in the inset. White arrows additionally highlight domain wall magnetization orientation. (c) Histogram corresponding to (a) showing the normalized counts (each dot representing counts after binning) of the angle α, defined as the angle between the DW magnetization vector $\boldsymbol{m}$ and the DW normal vector $\boldsymbol{n}$, as shown in the inset. A single peak at α=0° demonstrates left-handed Néel DW structure. (d) Histogram corresponding to (b) showing the normalized count of the angle χ, defined as the angle between the DW magnetization vector $\boldsymbol{m}$ and the domain magnetization $\boldsymbol{m_d}$, as shown in the inset; Peaks at χ=±90° and 270° indicate achiral Néel DW structure.

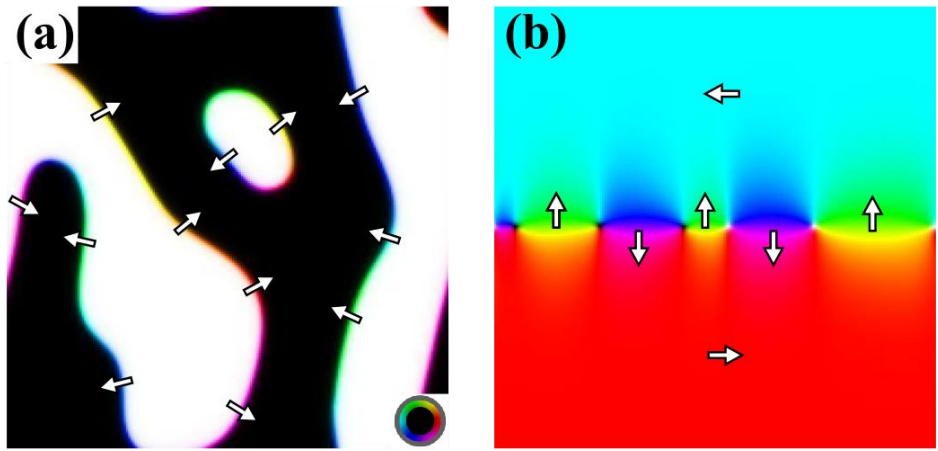

FIG. 3: Monte Carlo simulations for (a) out-of-plane magnetized and (b) in-plane magnetized thin film systems with equally strong DMI, showing chiral Néel domain walls in (a) and achiral Néel domain walls in (b). Local magnetization directions are rendered in color according to the color wheel in the inset. In addition, arrows highlight magnetization directions in domains and domain walls.

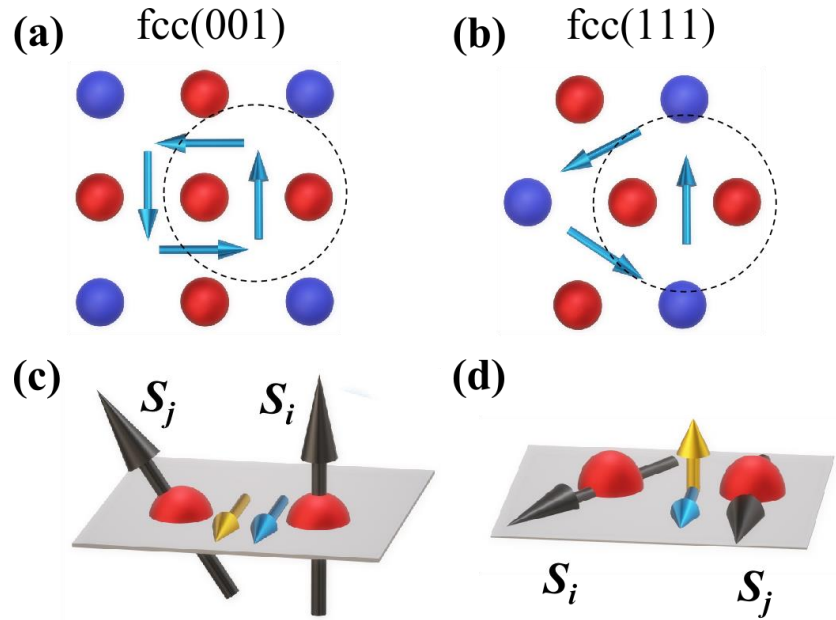

FIG. 4 (a,b) DMI vector (teal) for fcc(001) and fcc(111) respectively. Red spin sites denote selected nearest neighbors to a central site between which the DMI vector ($\mathbf{D}_{ij}$) is shown in teal. The dotted circles facilitate comparison between fcc(001) and fcc(111), showing the same relative orientation of the DMI vector with respect to neighboring spin sites. (c,d) Neighboring spin sites $\mathbf{S}_i$ and $\mathbf{S}_j$ (red) in right-handed chiral (c) out-of-plane and (d) in-plane magnetized Néel walls. The magnetization vector, DMI vector ($\mathbf{D}_{ij}$), and cross product ($\mathbf{S}_i \times \mathbf{S}_j$) are denoted by the black, teal, and yellow arrows respectively.

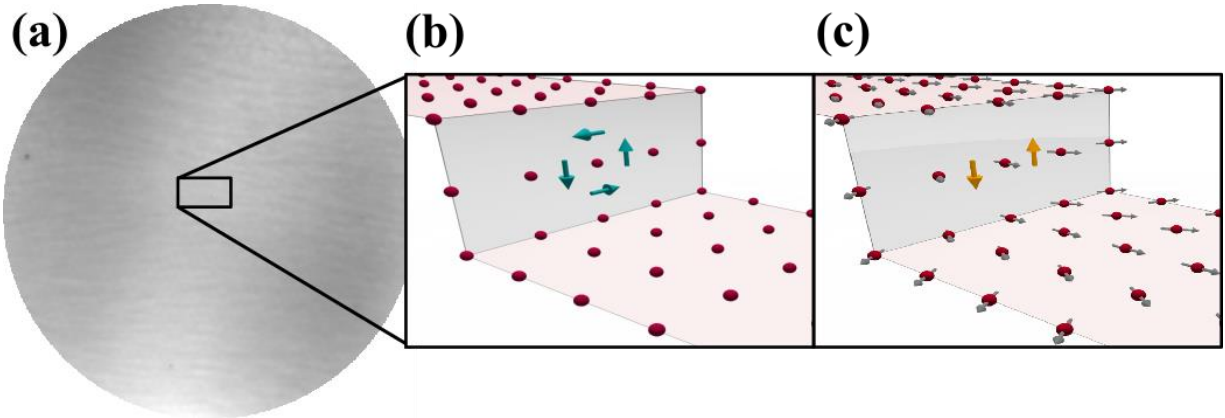

FIG. 5 (a) LEEM image showing atomic steps (dark lines) of Cu(001), over an 8 μm area [42]. (b) The possible DMI vector $\mathbf{D}_{ij}$ at an atomic double-step is represented by teal arrows. (c) The cross product $(\mathbf{S}_i \times \mathbf{S}_j)$ between neighboring spin sites at an atomic double-step (where $\mathbf{S}_i$ is always the central spin and $\mathbf{S}_j$ is always the neighbor), shown by yellow arrows, that will contribute to a nonzero $E_{\mathrm{DM}}$.